\documentstyle[prl,aps,multicol,epsfig,amsmath]{revtex}

\begin{document}
\draft
\tightenlines

\title{\bf Transition from Knudsen to molecular diffusion\\
in activity of absorbing irregular interfaces}

\author{J. S. Andrade Jr.$^{1,4}$, H. F. da Silva$^{1,2}$,
M. Baquil$^{1}$ and B. Sapoval$^{3,4}$}

\address{$^1$Departamento de F\'\i sica, Universidade Federal do Cear\'a,\\
60451-970 Fortaleza, Cear\'a, Brazil\\
$^2$Departamento de F\'{\i}sica, Universidade Federal do Maranh\~{a}o,\\
65080-040 S\~ao Luis, Maranh\~ao, Brazil\\
$^3$Laboratoire de Physique de la Mati\`ere Condens\'ee,  CNRS, Ecole 
Polytechnique
,\\91128 Palaiseau, France\\
$^4$Centre de Math\'ematiques et de leurs Applications, CNRS, 
\\Ecole Normale Sup\'erieure de Cachan,94235 Cachan, France}
\date{\today}
\maketitle
\begin{abstract}
We investigate through molecular dynamics the transition from Knudsen
to molecular diffusion transport towards $2d$ absorbing interfaces
with irregular geometry. Our results indicate that the length of the
active zone decreases continuously with density from the Knudsen to
the molecular diffusion regime. In the limit where molecular diffusion
dominates, we find that this length approaches a constant value of the
order of the system size, in agreement with theoretical predictions
for Laplacian transport in irregular geometries. Finally, we show that
all these features can be qualitatively described in terms of a simple
random-walk model of the diffusion process.
\end{abstract}


\begin{multicols}{2}
\narrowtext

The problem of Laplacian transport towards irregular surfaces
represents an important subject of research in many fields of
technological relevance including heterogeneous catalysis, heat
transfer and electrochemistry. In the particular case of catalysis,
the role of the local surface morphology at the pore level on the
global diffusion-reaction efficiency of the catalyst has been a
subject of great interest in the past years
\cite{Froment90,Thomas97,Gutfraind91,Coppens99,Sheintuch01}. 
Fractal concepts have been used to model complex surface roughness 
in the study of Knudsen diffusion in irregular reactive pores
\cite{Coppens95,Santra98} including catalyst supports and adsorbents
\cite{Avnir89}.

Another interesting point on the subject refers to the problem of the
nonuniform accessibility of active sites located along an irregular
reactive surface. If the system is diffusion-controlled, these
so-called {\it screening} effects may cause a dramatic reduction on
the reactivity of the catalyst surface, as compared to the reactivity
solely due to the intrinsic chemical reaction mechanism (i.e., the
activity of the nominal surface). The extensive research developed on
this field has been mainly devoted to the introduction, calculation
and application of the concept of {\it active zone} in the Laplacian
transport to and across irregular interfaces \cite{Sapoval94,Sapoval96}.  
For example, through the coarse-graining method proposed in
\cite{Sapoval94}, it is possible to determine the flux through an
arbitrarily irregular surface from its geometry alone, avoiding the
solution of the Laplace problem within a complex boundary domain. More
recently, it has been shown that this technique provides consistent
predictions for the activity of catalyst surfaces \cite{Sapoval01}. 

All these studies dealing with active zone in Laplacian fields rely on
the implicit assumption that molecular diffusion is the governing
mechanism of mass transport. Such an approximation, however, can only
be locally valid inside of the void space between the fins or extended
protrusions of an irregular surface if the mean free path of the
diffusing molecules is sufficiently smaller than the width of these
irregularities. For example, Knudsen diffusion may become the dominant
mechanism of mass transport determining the reactivity of the system
if the reagent is a diluted gas for which the collisions among
molecules are less frequent than the collisions between the molecules
and the catalytic surface \cite{Coppens95,Santra98}. The molecular
mean free path therefore constitutes a lower cut-off for the validity
of the molecular diffusion description. In a recent study \cite{Malek01}, 
it has been shown analytically and through numerical simulations that
surface roughness can have a significant effect on self-diffusion of
gases in nanoporous media in the Knudsen regime.

The aim of the present Letter is to study the transition in activity
of an irregular absorbing interface when the mechanism of mass
transport changes from Knudsen to molecular diffusion. Our approach is
to use a nonequilibrium molecular dynamics (NMD) technique in order to
simulate a nonuniform and steady-state profile of reagent
concentration between two active interfaces with an arbitrarily given
roughness. This resembles more closely the experimental conditions
frequently used in diffusion-reaction measurements. Here we adopt an
NMD method that has been originally proposed for the study of
self-diffusion in pure fluids \cite{Erpenbeck77}. The technique is
entirely based on the standard molecular dynamics (MD) at equilibrium,
but includes a special scheme to identify and exchange {\it labeled}
and {\it unlabeled} particles during the simulation.

The MD part of the simulation consists in a two-dimensional cell of
size $L_{x} l \times L_{y} l$ containing $N$ identical
particles that interact through the Lennard-Jones potential,
$\Phi(\Delta r_{ij})=4 \epsilon [(\sigma /\Delta r_{ij})^{12}-(\sigma
/\Delta r_{ij})^{6}]$, where $\Delta r_{ij}$ is the distance between
particles $i$ and $j$, $\epsilon$ is the minimum energy, $\sigma$ is
the zero of the potential, and we use $l/\sigma=5.86$. Periodic
boundary conditions are applied in both the $x$ and $y$
directions. Distance, energy and time are measured in units of
$\sigma$, $\epsilon$ and $(m {\sigma^2}/{\epsilon})^{1/2}$,
respectively, and the equations of motion are numerically integrated
using the Verlet algorithm \cite{Allen87}. In all simulations we
performed, the relative fluctuation around the average of the total
energy of the system has always been smaller than $10^{-4}$.

After thermalization, two identical fractal interfaces of size
$L_{y}l$ and perimeter $L_{p}l$ are symmetrically placed into the
system to simulate the roughness geometry of an absorbing material
(see Fig.~1). At this point, the non-equilibrium dynamics is put
forward through the following scheme: {\bf (1)} half of the particles
in the MD cell are randomly selected to carry a {\it label}, while the
other half are left {\it unlabeled}; {\bf (2)} every time a labeled
(unlabeled) particle crosses the interface at right (left) moving in
the $\vec{e_x}$ ($-\vec{e_x}$) direction it becomes unlabeled
(labeled), and {\bf (3)} when reinjected to the right (left) through
the periodic boundaries in the $x$ axis, an unlabeled (labeled)
particle becomes labeled (unlabeled). Although the particles are
indistinguishable in terms of their mass and interaction potential,
this labeling technique naturally builds concentration gradients for
both ``species'' that gradually develop to reach a desired non-uniform
steady-state. In Fig.~1 we show the resulting stationary profiles
along the $x$ coordinate of the number fractions $\theta_{l}\equiv
n_{l}/(n_{l}+ n_{u})$ and $\theta_{u}=1-\theta_{l}$, where $n_{l}$ and
$n_{u}$ are the number of labeled and unlabeled particles,
respectively, inside a vertical slice of fixed length in the
system. From this point on during the simulation, we keep updating at
each time step the number $n_{i}$ of particles being ``absorbed'' by
the element $i$ of the interface in order to compute its local mass
flux $q_{i}=n_{i}/\Delta t$, where $\Delta t$ is the elapsed time
after the steady-state has been established. Here we measure the
efficiency of the interface in terms of the active length $L_{a}$
defined as \cite{Sapoval}
\begin{equation}
L_{a}\equiv 1/\sum_{i=1}^{L_{p}}\phi_{i}^2~~~~
(1 \leq L_{a} \leq L_{p})~,
\label{eq1}
\end{equation}
where the sum is over the total number of interface elements $L_{p}$,
and $\phi_{i} \equiv q_{i}/\sum q_{j}$ is the normalized mass flux at
element $i$. From the definition (\ref{eq1}), $L_{a}=L_{p}$ indicates
a limiting state of equal partition of fluxes (${\phi_i}=1/L_{p}$,
$\forall i$) whereas $L_{a}$ of order $L_{y}$ should correspond to the
case of an strongly ``localized'' flux distribution.

Based on this NMD method, we performed simulations for 5 distinct
initial configurations of the dynamical system, different values of
the reduced temperature $T$, and reduced densities in the range $0.025
\leq \rho \leq 0.5$, corresponding to systems with $N=1250$ to $25000$
particles. The evolution with time of the active length for any NMD
realization reveals that, after a transient period, the system reaches
a stationary state characterized by an average value of $L_{a}$ that
is representative of the flux distribution. As shown in Fig.~2, the
active length decreases sharply with $\rho$ for low density systems at
$T=1.25$, up to a point where it remains constant at $L_{a} \approx
27$. The results from simulations performed at a higher temperature,
$T=3.33$, show that the behavior of the active length remains nearly
the same, at least within the range of densities considered here. 

The decrease with density of the active length $L_{a}$ reflects the
transition from Knudsen to molecular diffusion in the distribution of
activity at the interface. Because the mean free path of the particles
for small $\rho$ values is larger than the smaller length scale $l$ of
the irregular interface, the activity is highly sensitive to
geometrical constraints in the Knudsen regime. As shown in the inset
of Fig.~2, the difference in behavior of the pair-correlation function
for dilute and more dense fluids indicates that the simulated system
undergoes noticeable structural modifications as $\rho$ increases.

At higher densities, the invariant behavior of $L_{a}$ is a
consequence of molecular diffusion and can be explained in terms of an
interesting theorem given by Makarov \cite{Makarov85} to describe the
properties of Laplacian fields on two-dimensional interfaces of
arbitrary shape subjected to Dirichlet boundary conditions. The
theorem states that {\it the information dimension of the harmonic
measure on a singly connected interface in $d=2$ is exactly equal to
1}. In terms of activity, this means that, regardless the shape of the
interface, the total length $L_{a}$ of the region where most of the
activity takes place should be of the order of the size $L$ of the
cell under a dilation transformation (see Ref.~\cite{Sapoval} for a
detailed discussion of the active zone concept). Translating to our
diffusion cell, where square Koch trees of third generation are the
absorbing interfaces, the theorem of Makarov predicts that the value
of $L_{a}$ should be close to the size $L_{y}=27$, in good agreement
with the NMD limit obtained for high densities.

One can compute directly the length of the active zone for the
two-dimensional continuum Laplacian problem which represents
steady-state diffusion \cite{Evertsz92}. Numerical solutions of the
Laplace equation, ${\nabla}^{2} C=0$, for the concentration field
$C$ inside the diffusion cell are obtained here through numerical
discretization by means of finite-differences. For this, a constant
unitary concentration is imposed at the source line ($C_{0}=1$) and
Dirichlet boundary conditions ($C=0$) are assigned to each elementary
unit of the interface.  Due to the symmetry with respect to the source
line, only the concentration field in half of the domain needs to be
calculated. The calculated value for the active length of the purely
Laplacian cell is found to be $L_{a}=22.9$, compatible with
the prediction of the Makarov theorem, $L_{a} \approx L_{y}$.

Now we show that a simple random-walk model of the
diffusion-absorption process can provide a consistent qualitative
description of the behavior observed in the NMD simulations. Adopting
the same geometry, a particle is released from a random position in
the center line. The walker travels through the medium taking steps of
random directions, but constant length $\lambda$, till it crosses one
of the wall elements of the fractal interface and gets absorbed. The
flux at this element is then updated and the active length $L_{a}$ of
the interface recalculated. For a fixed value of the step length
$\lambda$, the simulation goes on with particles being sequentially
released and absorbed, till the active length reaches an average value
that is approximately constant. This value is usually obtained with
less than $10^5$ particles launched in the system. In Fig.~3 we show
the dependence on the parameter $\xi \equiv (\sigma/\lambda)$ of the
average $L_{a}$ computed for the third generation of the square Koch
tree. For a two-dimensional gas, $\lambda$ can be interpreted as the
mean free path, which is inversely proportional to the surface density
of the system, $\lambda \propto 1/\rho$. Similarly to the NMD
simulations, two distinct regimes of activity can be clearly
identified and directly related to the different governing mechanisms
of mass transport, namely, Knudsen and molecular diffusion. At low
values of $\xi$, the sharp decrease of $L_{a}$ reflects the strong
influence on the mass transport process of the fractal geometry of the
interface. The semi-log plot shown in the inset of Fig.~3 indicates
that this decay in activity characterizing the Knudsen regime is
approximately logarithmic in shape.

At sufficiently large values of $\xi$, the length $L_{a}$ reaches a
plateau of minimum activity that is practically coincident with the
value of the active length found for the Laplacian cell, $L_{a}=22.9$
(dashed line at the bottom in Fig.~3). It is important to show that
these two approaches to the problem provide consistent results for
denser systems, even at the local scale of the interface
geometry. Indeed, as displayed in Fig.~4, the normalized fluxes at
each wall element generated from the continuum and random-walk methods
are almost indistinguishable. Compared to the lower limit of the
random-walk model, $L_{a}\approx 24$, the higher value found for the
active length with the NMD technique, $L_{a}\approx 27$, can be
explained in terms of the structural features of the simulated
fluid. As shown in the inset of Fig.~2, the role of the attractive
part of the interaction potential is to induce a smooth peak of
short-range correlation that characterizes the presence of small
particle clusters in the fluid. This geometrical aspect of the NMD
system tends to increase the measure of the active length of the
absorbing interface.

In conclusion, we have shown through molecular dynamics simulations
that the active fraction of an irregular absorbing interface should be
sensitive to: (i) its geometrical details; (ii) the governing
mechanism of transport, and (iii) the structural aspects of the
diffusing fluid. In particular, the active length for absorption of
molecular diffusing fluids is found to be very close to that of a
purely Laplacian system. These observations may lead to new guidelines
to the problem of diffusion and absorption on arbitrarily irregular
interfaces. Furthermore, we have proposed a very simple random-walk
model that incorporates the basic features of the diffusion-absorption
process and is capable to describe, at least semi-quantitatively, the
behavior of the active length for different diffusion regimes. This
model provides substantial insight on the effect of the diffusion
mechanism on the interface activity and has the virtue of being
computationally cheap. Finally, the approach introduced here is
flexible enough to represent specific characteristics of irregular
interfaces as well as other types of ``absorption'' mechanisms (e.g.,
finite-rate chemical reactions) limited by diffusion transport. As a
consequence, the implications of this study are certainly relevant for
the analysis and interpretation of diffusion and reaction processes
occurring in a large variety of catalyst materials. The modeling
techniques we adopted should also be useful as design tools to choose
a suitable catalytic interface for a given reaction-diffusion system.

We thank CNPq, CAPES, COFECUB and FUNCAP for support. The Centre de
Math\'ematiques et de leurs Applications and the Laboratoire de
Physique de la Mati\`{e}re Condens\'{e}e are ``Unit\'{e} Mixte de
Recherches du Centre National de la Recherche Scientifique'' no. 8536
and 7643.

\begin{figure}
\includegraphics[width=8.0cm]{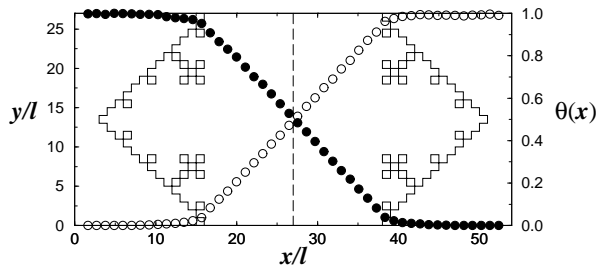}
\caption{Schematic representation of the NMD diffusion cell. The
absorbing interfaces are square Koch trees. Also shown in this figure
is the dependence of the local number fraction $\theta$ of labeled
(full circles) and unlabeled (empty circles) particles on the position
along the $x$ direction in the cell. In the case of the random-walk
model, particles are released from random $y$ positions at the dashed
line in the center.}
\end{figure}

\begin{figure}
\includegraphics[width=8.0cm]{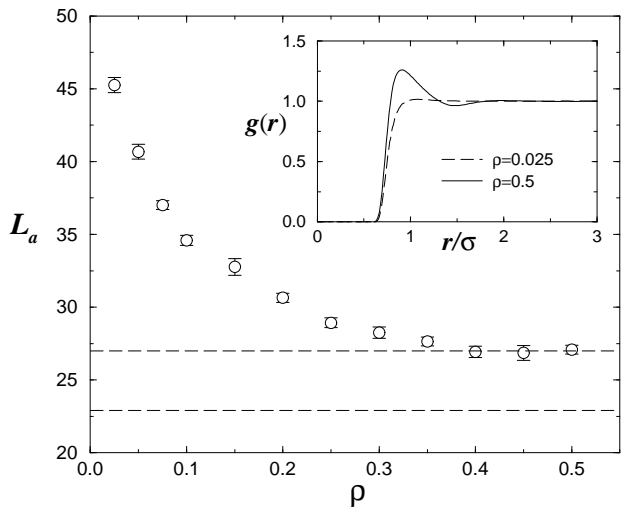}
\caption{Dependence of the active length $L_{a}$ on the reduced
density of the NMD cell for a fixed temperature, $T=1.25$. The
average values with error bars refer to simulations with $5$
different realizations of the NMD process. The horizontal dashed line
at the top corresponds to the system size, $L_{y}=27$, while the one
at the bottom indicates the value of the active length obtained from
the simulation with the Laplacian cell, $L_{a}=22.9$. The inset shows
the pair-correlation function calculated at the same temperature for a
dilute ($\rho=0.025$) and a more dense ($\rho=0.5$) fluid.}
\end{figure}

\begin{figure}
\includegraphics[width=8.0cm]{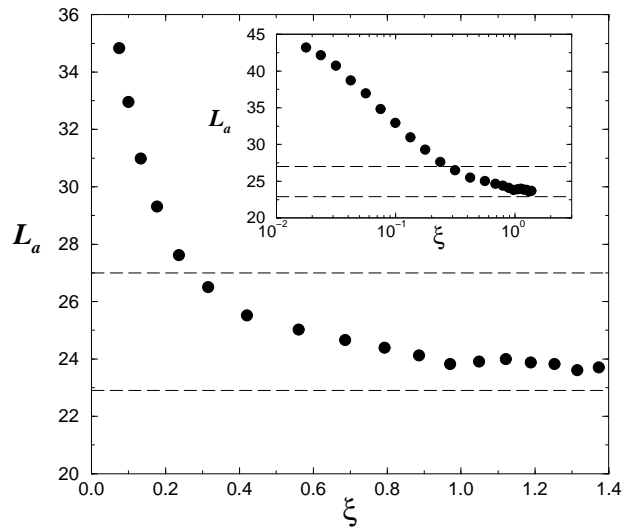}
\caption{Dependence on the random-walk parameter $\xi$ of the active 
length $L_{a}$ of the fractal interface for the random-walk model. The
horizontal dashed line at the top corresponds to the system size,
$L_{y}=27$, while the one at the bottom gives the active length of the
Laplacian cell, $L_{a}=22.9$. The semi-log plot in the inset shows
that $L_{a}$ decays approximately logarithmically with $\xi$ in the
regime of Knudsen diffusion.}
\end{figure}

\begin{figure}
\includegraphics[width=8.0cm]{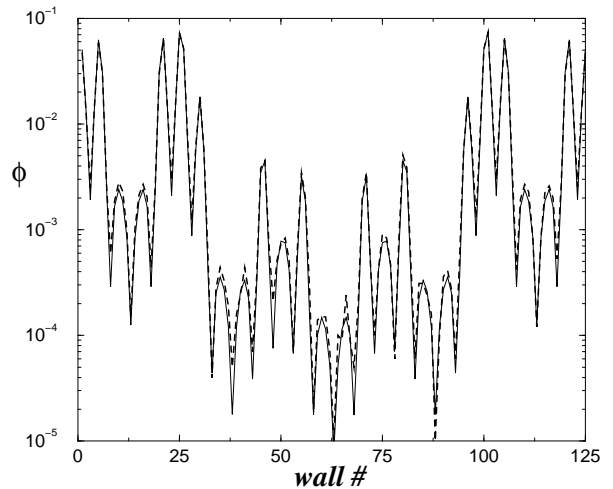}
\caption{Distribution of the logarithm of the normalized fluxes
crossing the wall elements along the absorbing fractal interface.
The dashed line corresponds to a random-walk simulation performed with
$\xi=0.63$, and the solid line is the distribution resulting
from the numerical simulation with a Laplacian cell.}
\end{figure}

\end{multicols}
\end{document}